\documentclass[a4paper]{jpconf}
\usepackage{graphicx}
\usepackage{clshan-math,clshan-dm-dd}
\def \lsim {\:\raisebox{-0.7ex}{$\stackrel{\textstyle<}{\sim}$}\:}
\def \gsim {\:\raisebox{-0.7ex}{$\stackrel{\textstyle>}{\sim}$}\:}
\begin{document}
\title{Background effects              \vspace{-0.15cm}
       on reconstructed WIMP couplings \hspace*{-0.3cm}}
\author{Chung-Lin Shan}
\address{\it Institute of Physics, Academia Sinica    \\
             No.~128, Sec.~2, Academia Road, Nankang,
             Taipei 11529, Taiwan, R.O.C.}
\ead{clshan@phys.sinica.edu.tw}
\begin{abstract}
 In this talk,
 I presented effects of
 small, but non--negligible unrejected background events
 on the determinations of WIMP couplings/cross sections.
\end{abstract}
\section{Introduction}
 Residue background events
 which pass all discrimination criteria and
 then mix with other real signals in data sets
 are one of the most important issues in all underground experiments.
 Hence,
 as a more realistic study,
 we reexame our model--independent data analysis methods
 for determining (ratios between) different couplings/cross sections of
 Weakly Interacting Massive Particles (WIMPs) on nucleons
 in direct Dark Matter detection experiments
 \cite{DMDDfp2, DMDDranap}
 by taking into account
 small fractions of unrejected background events
 \cite{DMDDbg-fp2, DMDDbg-ranap}.
\section{Background effects on reconstructed WIMP--nucleon couplings/cross sections}
 In this section
 I present simulation results
 of the reconstructed (ratios between)
 different WIMP--nucleon couplings/cross sections
 with mixed data sets
 from WIMP--induced and background events.
 2 (3) $\times$ 5,000 experiments have been simulated
 by the Monte Carlo method.
 Each experiment contains 50 total events
 on average.
 The experimental threshold energies of all experiments
 have been assumed to be negligible
 and the maximal cut--off energies
 are set the same as 100 keV.
 Note here that
 both signal and background events
 are treated as WIMP signals in the analyses.

\subsection{Reconstructed $|f_{\rm p}|^2$}
\begin{figure}[t!]
\begin{center}
\includegraphics[width=9cm]{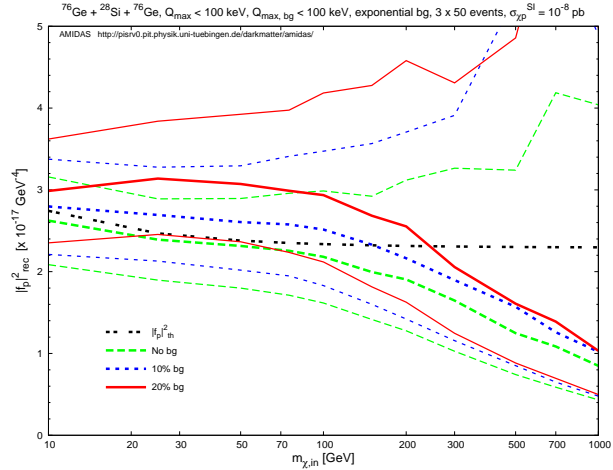} \\
\vspace{-0.5cm}
\end{center}
\caption{
 The reconstructed squared SI WIMP--nucleon couplings $|f_{\rm p}|^2$
 and the lower and upper bounds of
 their 1$\sigma$ statistical uncertainties
 as functions of the input WIMP mass
 reconstructed with a $\rmXA{Ge}{76}$ target.
 Another {\em independent} data set of $\rmXA{Ge}{76}$
 and one of $\rmXA{Si}{28}$
 have been used for determining the needed WIMP mass.
 The double--dotted black curve
 is the theoretical value of $|f_{\rm p}|^2$
 corresponding to the fixed SI WIMP--nucleon cross section
 $\sigmapSI = 10^{-8}$ pb.
 The background ratios shown here
 are no background (dashed green),
 10\% (long--dotted blue)
 and 20\% (solid red)
 (plot from Ref.~\cite{DMDDbg-fp2}).
}
\label{fig:fp2-Ge-SiGe-ex-000-100-050}
\end{figure}

 In Fig.~\ref{fig:fp2-Ge-SiGe-ex-000-100-050}
 I show the reconstructed squared SI WIMP--nucleon couplings $|f_{\rm p}|^2$
 as functions of the input WIMP mass.
 The SI WIMP--nucleon cross section
 is set as $10^{-8}$ pb,
 the commonly used value for the local WIMP density
 $\rho_0 = 0.3~{\rm GeV/cm^3}$
 has been used for both data generating and analyzing.
 A $\rmXA{Ge}{76}$ target has been used
 for reconstructing $|f_{\rm p}|^2$,
 whereas another {\em independent} data set of $\rmXA{Ge}{76}$
 and one of $\rmXA{Si}{28}$
 have been used for determining the needed WIMP mass.
 The background ratios shown here
 are no background (dashed green),
 10\% (long--dotted blue)
 and 20\% (solid red).

 It can be seen here that,
 not surprisingly,
 due to extra unexpected background events,
 the larger the background ratio in the analyzed data set,
 the more strongly overestimated
 the reconstructed SI WIMP--nucleon coupling
 for all input WIMP masses.
%
 Moreover,
 for a heavy target, e.g., $\rmXA{Ge}{76}$,
 it has been found that
 this overestimate could be at the strongest
 for WIMP masses between 30 GeV and 100 GeV,
 once the background ratio rises to $\gsim$ 20\%
 \cite{DMDDbg-fp2}.
 Nevertheless,
 Fig.~\ref{fig:fp2-Ge-SiGe-ex-000-100-050}
 shows that
 one could in principle estimate the SI WIMP--nucleon coupling
 with an uncertainty of a factor $\lsim$ 2
 by using data sets
 with maximal 20\% background events.

%
%
\subsection{Reconstructed $\armn / \armp$}

\begin{figure}[t!]
\begin{center}
\includegraphics[width=6.75cm]{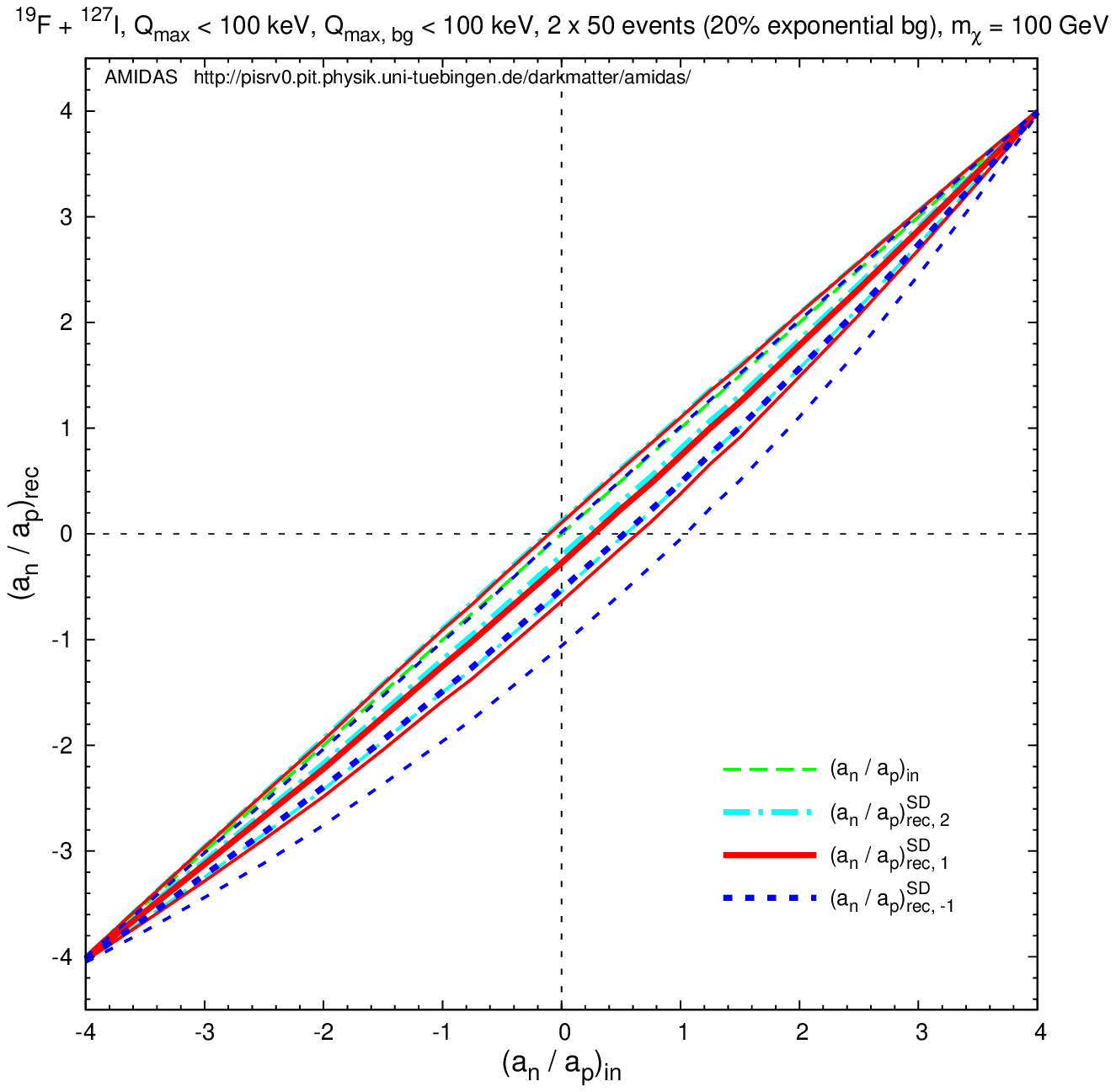}         \hspace{0.75cm}
\includegraphics[width=6.75cm]{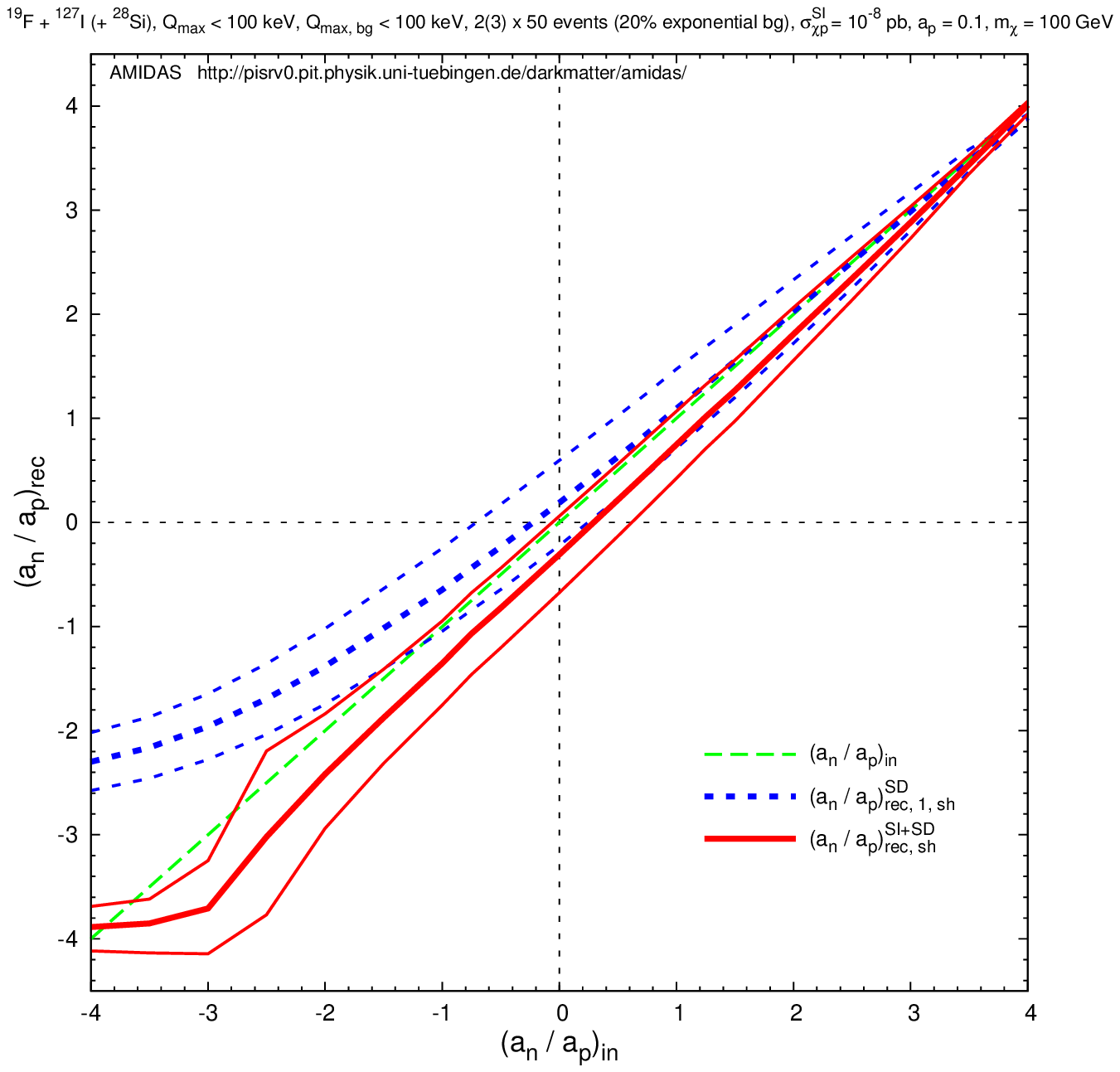} \\
\vspace{-0.5cm}
\end{center}
\caption{
 The reconstructed $\armn / \armp$ ratios
 and the lower and upper bounds of
 their 1$\sigma$ statistical uncertainties
 as functions of the input $\armn / \armp$ ratio.
 Left:
 a SD--dominant WIMP--nucleus interaction
 with $n = -1$ (dashed blue), 1 (solid red),
 and 2 (dash--dotted cyan);
 right:
 adding a non--zero SI WIMP--nucleus interaction,
 analyzed with (solid red) and without (dashed blue, $n = 1$)
 taking into account this term.
 The mass of incident WIMPs
 has been set as \mbox{$\mchi = 100$ GeV}.
 The input SI WIMP--nucleon cross section
 and the input SD WIMP--proton coupling
 have been set as \mbox{$\sigmapSI = 10^{-8}$ pb} and $\armp = 0.1$,
 respectively.
 The background ratio in the analyzed data sets
 is 20\%
 (plots from Ref.~\cite{DMDDbg-ranap}).
}
\label{fig:ranapSISD-08-ex}
\end{figure}

 The left frame of Figs.~\ref{fig:ranapSISD-08-ex}
 shows the reconstructed $\armn / \armp$ ratios
 with $n = -1$ (dashed blue), 1 (solid red),
 and 2 (dash--dotted cyan)
 as functions of the input $\armn / \armp$ ratio
 for the case of a dominant SD WIMP--nucleus interaction.
 A combination of \mbox{$\rmXA{F}{19}$ + $\rmXA{I}{127}$} targets
 has been used.
 The background ratio in the analyzed data sets
 is 20\%.
 The mass of incident WIMPs
 has been set as \mbox{$\mchi = 100$ GeV}.

 It can be seen that,
 due to the non--negligible background events,
 the reconstructed $\armn / \armp$ ratios
 (for this \mbox{$\rmXA{F}{19}$ + $\rmXA{I}{127}$} target combination)
 become now underestimated;
 in fact,
 the larger the background ratio
 the larger this systematic deviation of
 the reconstructed $\armn / \armp$ ratios
 \cite{DMDDbg-ranap}.
 Meanwhile,
 for the same data sets,
 the larger the $n$ value
 (or, equivalently,
  the larger the used moment of the WIMP velocity distribution),
 the smaller this systematic deviation
 \cite{DMDDbg-ranap}.
 This (in)compatibility
 between the reconstructed $\armn / \armp$ ratios with different $n$
 could thus offer us a simple check
 for the purity/availability of our data sets.

 On the other hand,
 in the right frame of Figs.~\ref{fig:ranapSISD-08-ex}
 I show the reconstructed $\armn / \armp$ ratios
 for the case of a general combination of both SI and SD WIMP interactions.
 Here we analyzed the same data sets
 with (solid red) and without (dashed blue, $n = 1$)
 considering the extra SI interaction term.

 It can be seen that
 the (in)compatibility
 between the reconstructed $\armn / \armp$ ratios
 under different assumptions about the WIMP--nucleus interactions
 becomes larger.
%
 Moreover
 it has been found that,
 firstly,
 with an increased background ratio
 the systematic deviations and the statistical uncertainties
 of the reconstructed $\armn / \armp$ ratio
 for considering the general combination of
 the SI and SD WIMP interactions (solid red)
 grow only (very) slightly
 \cite{DMDDbg-ranap}.
 Secondly,
 once residue background events
 exist regularly between the experimental
 minimal and maximal cut--off energies
 or (even better) (mostly) in high energy ranges,
 one can in principle estimate
 the ratio between two SD WIMP--nucleon couplings
 (pretty) precisely
 {\em without} worrying about
 the non--negligible backgrounds
 \cite{DMDDbg-ranap}.

 Moreover,
 by considering different input WIMP masses,
 a WIMP--mass independence
 of the reconstructed $\armn / \armp$ ratios
 as well as of their statistical uncertainties
 with even non--negligible background events
 has been observed
 \cite{DMDDbg-ranap}.
 With data sets of $\lsim~20\%$ background ratios,
 the reconstructed 1$\sigma$ statistical uncertainty intervals
 could in principle always cover
 the input (true) $\armn / \armp$ ratios
 pretty well
 for WIMP masses $\gsim$ 25 GeV
 \cite{DMDDbg-ranap}.

\subsection{Reconstructed $\sigma_{\chi ({\rm p, n})}^{\rm SD} / \sigmapSI$}
\begin{figure}[t!]
\begin{center}
\includegraphics[width=6.75cm]{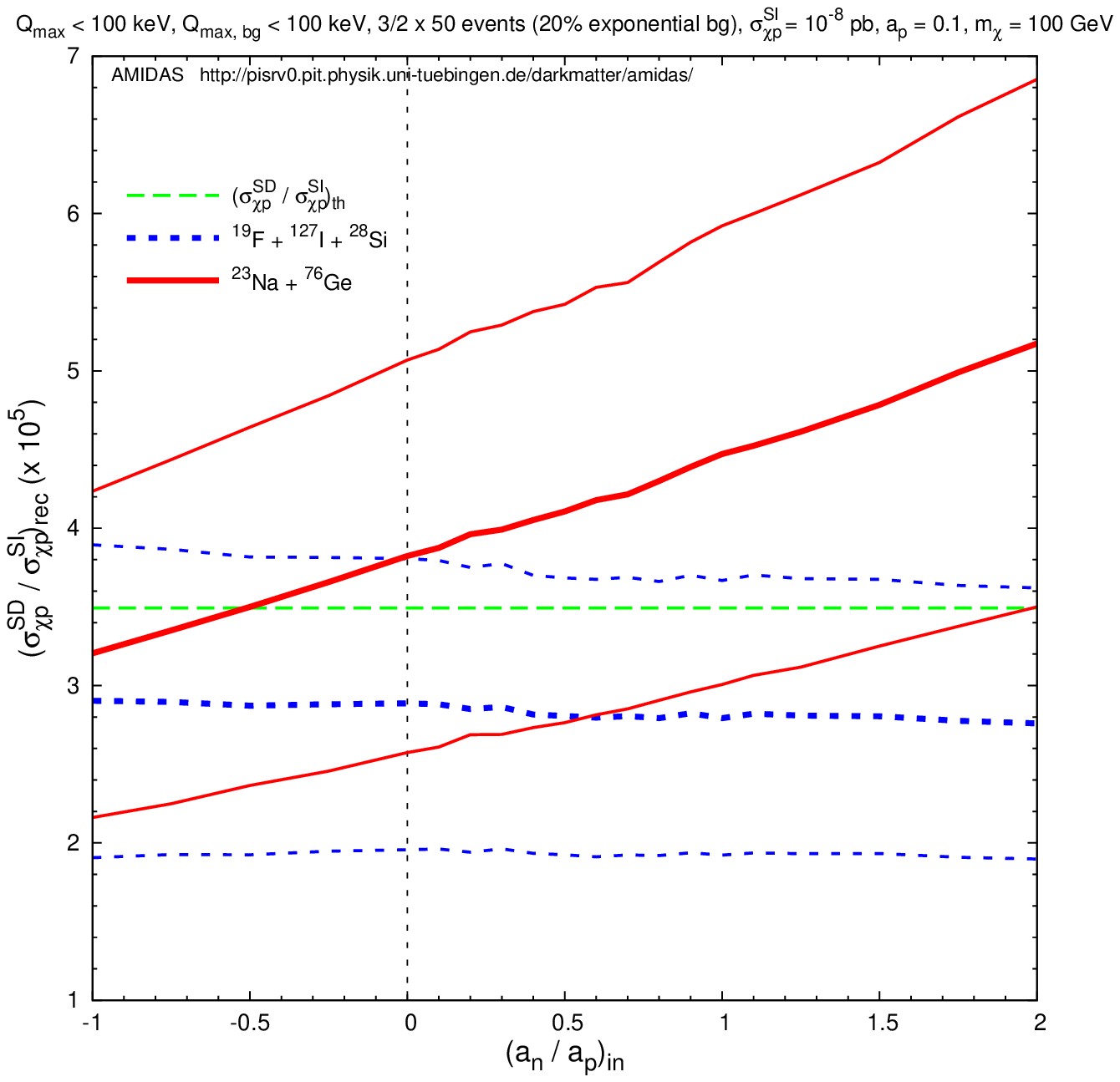} \hspace{0.75cm}
\includegraphics[width=6.75cm]{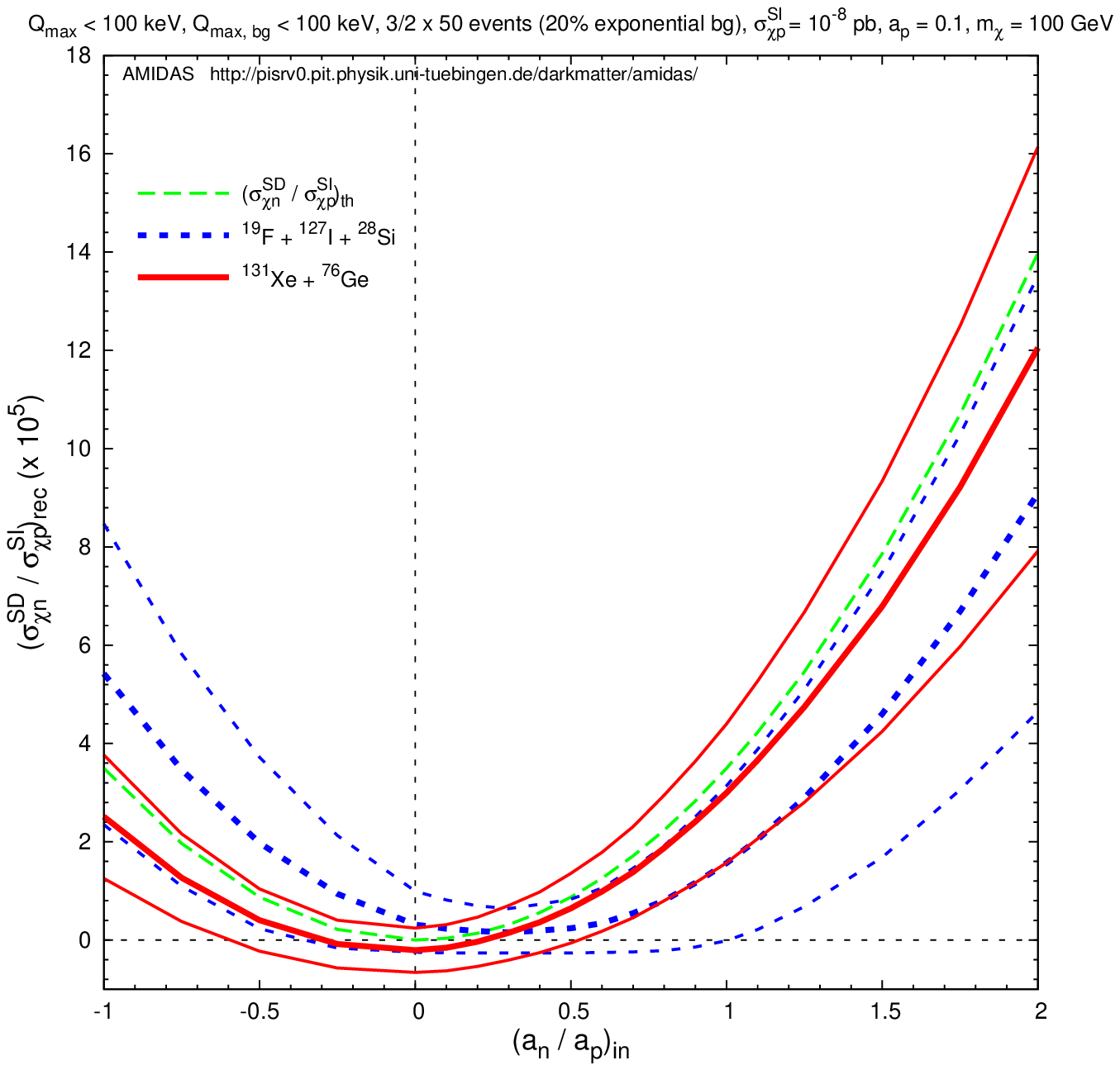} \\
\vspace{-0.5cm}
\end{center}
\caption{
 The reconstructed $\sigmapSD / \sigmapSI$ (left)
 and $\sigmanSD / \sigmapSI$ (right) ratios
 and the lower and upper bounds of
 their 1$\sigma$ statistical uncertainties
 as functions of the input $\armn / \armp$ ratio.
 The dashed blue curves indicate the ratios
 reconstructed with $\rm F + I + Si$,
 whereas the solid red curves indicate the ratios
 with
 $\rmXA{Ge}{76}$
 and combined with
 $\rmXA{Na}{23}$ (left)
 and $\rmXA{Xe}{131}$ (right).
 Parameters are as
 in the right frame of Figs.~\ref{fig:ranapSISD-08-ex}
 (plots from Ref.~\cite{DMDDbg-ranap}).
}
\label{fig:rsigmaSDSI-08-ranap-ex}
\end{figure}

 Figs.~\ref{fig:rsigmaSDSI-08-ranap-ex}
 show the reconstructed $\sigmapSD / \sigmapSI$ (left)
 and $\sigmanSD / \sigmapSI$ (right) ratios
 as functions of the input $\armn / \armp$ ratio.
 It can be seen that,
 interestingly,
 while the $\sigmapSD / \sigmapSI$ ratios
 reconstructed with $\rm F + I + Si$ targets
 become more and more strongly underestimated
 with an increased background ratio,
 those reconstructed with $\rm Na + Ge$ targets
 become in contrast more and more strongly overestimated
 \cite{DMDDbg-ranap}.
 Meanwhile,
 more interestingly,
 while the $\sigmanSD / \sigmapSI$ ratios
 reconstructed with $\rm F + I + Si$ targets
 become more and more strongly underestimated
 with an increased background ratio
 for all input $\armn / \armp$ values,
 those reconstructed with $\rm Xe + Ge$ targets
 become more and more strongly {\em underestimated}
 for $\armn / \armp~\gsim~0$
 and more and more strongly {\em overestimated}
 for $\armn / \armp~\lsim~0$
 \cite{DMDDbg-ranap}.
 Nevertheless,
 with data sets of $\lsim~20\%$ background ratios,
 the reconstructed 1$\sigma$ statistical uncertainty intervals
 could in principle always cover
 the input (true) $\sigma_{\chi ({\rm p, n})}^{\rm SD} / \sigmapSI$ ratios
 pretty well
 \cite{DMDDbg-ranap}.

\section{Conclusions}
 In this article
 we reexamine the model--independent
 data analysis methods
 introduced in Refs.~\cite{DMDDfp2, DMDDranap}
 for determining (ratios between) different WIMP--nucleon couplings/cross sections
 from measured recoil energies of direct detection experiments directly
 by taking into account small fractions of residue background events.

 Our simulations show that,
 due to extra unexpected background events,
 the SI WIMP--nucleon coupling
 would be overestimated;
 nevertheless,
 the maximal acceptable background ratio
 in the analyzed data set
 of ${\cal O}(50)$ total events
 could be $\sim$ 20\%.

 On the other hand,
 the maximal acceptable background ratio
 for determining the ratio of the SD WIMP coupling
 on neutrons to that on protons
 is also $\sim$ 20\%.
 However,
 the larger the relative strength between
 the SD WIMP--nucleus interaction to the SI one,
 the smaller the systematic deviations
 as well as the statistical uncertainties.
 Moreover,
 it has also been found that,
 by taking different assumptions about the relative strength
 between the SI and SD WIMP--nucleus interactions,
 and/or using different moments of
 the one--dimensional WIMP velocity distribution,
 there would be an (in)compatibility
 between different reconstructed $\armn / \armp$ ratios,
 which could allow us to check
 the purity/availability of the analyzed data sets.

 Furthermore,
 we found also that
 only background events in the lowest energy ranges
 could affect the reconstructions (significantly);
 those in high energy ranges
 would almost not change the reconstructed ratios
 or only very slightly.

\section*{Acknowledgments}
%
 The author appreciate
 the Physikalisches Institut der Universit\"at T\"ubingen
 for the technical support of the computational work
 presented in this article.
 This work
 was partially supported
 by the National Science Council of R.O.C.~%
 under contract no.~NSC-99-2811-M-006-031.
\section*{References}
\end{document}